\documentstyle[12pt,aaspp4]{article}

\newcommand{\hmpc}{\;h^{-1}\rm{Mpc}}
\newcommand{\kms}{\rm{\;km\,s^{-1}}}
\newcommand{\order}{\!\sim\!}
\newcommand{\lp}{\left(}
\newcommand{\rp}{\right)}

\newcommand{\lb}{\left|}
\newcommand{\rb}{\right|}
\newcommand{\avg}[1]{\langle #1\rangle}

\def\epsilon{\varepsilon}

\newcommand{\CNOCI}{CNOC}
\newcommand{\CNOCII}{CNOC2}

\newcommand{\etal}{et al.}
\newcommand{\eg}{e.g.}
\newcommand{\vs}{vs}

\newcommand{\eqref}[1]{(\ref{eq:#1})}

\slugcomment{Submitted to {\it The Astrophysical Journal}}

\begin{document}

\newcommand{\zmin}{0.21}
\newcommand{\zmax}{0.53}
\newcommand{\dra}{27}
\newcommand{\ddec}{8}
\newcommand{\mmin}{17.0}
\newcommand{\mmax}{21.7}
\newcommand{\zmed}{0.36}
\newcommand{\nphot}{404}
\newcommand{\nclstr}{182}
\newcommand{\nfield}{144}
\newcommand{\ndiff}{38}
\newcommand{\gamfrc}{1.7}
\newcommand{\delfrc}{0.7}
\newcommand{\roloc}{5.1\hmpc}
\newcommand{\Jfac}{4.14}
\newcommand{\rof}{2.1\hmpc}
\newcommand{\rofe}{2.1^{+0.6}_{-0.3}\hmpc}
\newcommand{\efe}{1.8^{+0.9}_{-1.5}}
\newcommand{\sigf}{770\kms}
\newcommand{\sigfl}{840\kms}
\newcommand{\awpe}{1.3^{+0.4}_{-0.1}}
\newcommand{\awce}{1.2^{+1.2}_{-0.0}}
\newcommand{\roo}{2.5\hmpc}
\newcommand{\rooe}{2.5^{+0.7}_{-0.4}\hmpc}
\newcommand{\eoe}{0.8^{+1.0}_{-1.3}}
\newcommand{\sigo}{400\kms}
\newcommand{\sigol}{430\kms}
\newcommand{\xirof}{3.9\hmpc}
\newcommand{\xirofe}{3.9^{+0.7}_{-0.6}\hmpc}
\newcommand{\xiroo}{3.7\hmpc}
\newcommand{\xirooe}{3.7^{+0.7}_{-0.5}\hmpc}

\title{The Two-Point Correlation Function\\at Redshift $\slantfrac{{\bf{1}}}{{\bf{3}}}$}
\author{
C. W. Shepherd\altaffilmark{1}
R. G. Carlberg\altaffilmark{1,2}
H. K. C. Yee\altaffilmark{1,2}\and
E. Ellingson\altaffilmark{2,3}}
\altaffiltext{1}{Department of Astronomy, University of Toronto, Toronto,
Ont., Canada M5S 3H8}
\altaffiltext{2}{Guest observer, Canada-France-Hawaii Telescope,
operated jointly by NRC of Canada, CNRS of France, and the University of
Hawaii}
\altaffiltext{3}{C.A.S.A., University of Colorado, Campus Box 389, Boulder,
CO 80309-0389}

\begin{abstract}
We present the results of a study of the two-point correlation function
	for a sample of field galaxies taken from the \CNOCI\ cluster survey.
The sample consists of \nfield\ galaxies within a contiguous region of space
	subtending 225 square arcminutes.
The objects have $r$-band magnitudes $\mmin\le r\le\mmax$ and redshifts
	$\zmin\le z\le\zmax$.
The median redshift of the sample is $\bar z=\zmed$.
The real space correlation function is found to be consistent with a
	power law $\xi(r)=(\slantfrac{r}{{r_0}})^{-\gamfrc}$ with
	$r_0=\rofe$ ($\Omega_0=1$), or $r_0=\rooe$ ($\Omega_0=0.2$).
Uncertainties are estimated using the bias-corrected bootstrap resampling
	method, with 300 resamplings.
This low correlation length implies strong evolution since $z\order\zmed$ has
	occurred in either the correlation function or the luminosity function;
	if the observed correlation function is modeled as
	$\xi(r,z)=\xi(r,0)(1+z)^{-(3+\epsilon)}$ with
	$\xi(r,0)=(\slantfrac{r}{{\roloc}})^{-\gamfrc}$, then $\epsilon=\eoe$.
Comparison of the redshift space and real space correlation function
	indicates that the one-dimensional pairwise peculiar velocity
	dispersion $\sigma$ at $z\approx\zmed$ is weakly inconsistent with
	$\sigf$, the value predicted by the Cosmic Virial Theorem if
	$\Omega_0=1$.
The observed correlation function is, however, consistent with $\sigma=\sigo$, 
	the value expected if $\Omega_0=0.2$.
\end{abstract}


\section{Introduction}
\label{s:intro}

The study of the statistics of galaxy clustering has yielded important
	information about the large-scale structure of the universe and about the
	environment of galaxies.
One of the most useful statistics employed in this study is the two-point
	correlation function.
This statistic quantifies the clustering of galaxies, and is directly related
	to the power spectrum of density fluctuations in the galaxy distribution.
Determining the evolution of the correlation function is therefore essential
	for an understanding of cosmological structure formation.

The two-point correlation function has been extensively studied at low
	redshifts; \cite{Efsurv}\ gives a summary of redshift surveys that have
	been used for correlation analysis.
Observations of the correlation function at the present epoch indicate that it
	is well described by a power law
\begin{equation}
	\xi(r) = \lp r\over r_0\rp^{-\gamma}
\label{eq:plawdef}
\end{equation}
	at scales $r\lesssim10\hmpc$.
Here, and throughout this paper, all separations are given in physical, as
	opposed to comoving, coordinates unless otherwise stated; the Hubble
	parameter is taken to be $H_0=100h\kms Mpc^{-1}$.  
Typical results from optical surveys of nearby galaxies include
	$r_0=5.4\pm0.3\hmpc$, $\gamma=1.77\pm0.04$ from the Center for Astrophysics
	(CfA) survey (Davis \& Peebles 1983) and $r_0=5.1\pm0.2\hmpc$,
	$\gamma=1.71\pm0.05$ from the Stromlo-APM survey (Loveday \etal\ 1995).
Values for $r_0$ and $\gamma$ for local samples are typically taken to be
	$5.4\hmpc$ and $1.8$, respectively, based on the CfA observations.
For the purposes of comparing our results to those from local samples, however,
	we adopt $r_0=5.1\hmpc$ and $\gamma=1.7$, principally because of the
	fainter magnitude limits used in the Stromlo-APM survey.

The principal aim of this investigation is to examine the evolution of the
	correlation function.
A useful empirical model for this evolution, introduced by \cite{KnS}\ is
\begin{equation}
	\xi(r,z) = \xi(r,0) \lp1+z\rp^{-(3+\epsilon)}\;.
\label{eq:ev}
\end{equation}
In this model, $\epsilon=\gamma-3$ corresponds to clustering fixed in comoving
	coordinates, as seen in biased Cold Dark Matter (CDM) simulations
	(Carlberg  1991).
If $\epsilon=0$, clustering is stable in physical coordinates.
\cite{CnC}\ find $\epsilon\order0$ for an Open CDM (OCDM) initial power
	spectrum with $\Omega_0=0.2$.

To determine $\epsilon$, one must compute the correlation function from data
	from an earlier epoch.
The most common method is to compute the angular correlation function from the
	angular galaxy distribution (\eg\ Koo \& Szalay 1984, Efstathiou \etal\
	1991 and Infante \& Pritchet 1995).
This method does not require that the distances to the objects in the sample
	be known; rather, the clustering is observed as a two-dimensional
	projection of the three-dimensional clustering over a wide range of
	redshifts.
In order to estimate the amount of clustering evolution present in the data,
	therefore, models for both the present-day correlation function and the
	redshift distribution of objects in the sample must be employed.
Both Efstathiou \etal\ (1991) and Infante \& Pritchet (1995) find
	$\epsilon>0$ if $\Omega_0=1$ and there is no evolution in the luminosity
	function.

In order to produce a reliable estimate of the spatial two-point correlation
	function, redshifts for a large number of objects must be obtained.
Pencil-beam surveys using multi-object spectroscopy (MOS), such as the Autofib
	survey (Cole \etal\ 1994) and the Canada-France Redshift Survey (CFRS)
	(Le F\`evre \etal\ 1996) are the only surveys to date which contain a
	sufficient number of objects at intermediate redshifts.
Le F\`evre \etal\ (1996) find $\epsilon\order0-2$ in the CFRS survey,
	consistent with the evolution observed in angular surveys.

In principle, data from pencil-beam surveys may be used to estimate the
	redshift space correlation function.
The required velocity accuracy is, however, quite high.
\cite{CfA} find the one-dimensional pairwise peculiar velocity dispersion in
	the CfA survey to be $\sigma=340\pm40\kms$ at a separation of $1\hmpc$.
If the velocity errors in a survey are comparable to
	$\slantfrac{\sigma}{{\sqrt{2}}}$, computation of a reliable
	redshift space correlation function from that survey is impossible.
Cole \etal\ (1994) find a comoving correlation length $r_0=6.5\pm0.4\hmpc$
	based on the redshift space correlation function for the Autofib survey.
The median redshift of this survey is $\bar z=0.16$;
	taking $r_0(0.16)=\slantfrac{{6.5}}{{1.16}}\hmpc$ and $r_0(0)=5.1\hmpc$ in
	equation \eqref{ev} yields $\epsilon\approx-4$, at odds with previous
	angular correlation results.

We present here estimates of the real space and redshift space two-point
	correlation function for a sample of field galaxies taken from the
	Canadian Network for Observational Cosmology (CNOC) cluster survey.
The data are described in Section \ref{s:data}.
The techniques used for computing the real space two-point correlation
	function and its evolution, along with the results for this survey, are
	presented in Section \ref{s:analysis}.
Also described there is our investigation of the redshift space two-point
	correlation function and the heretofore unknown pairwise peculiar velocity
	dispersion at $z\order\zmed$.
Our conclusions are given in Section \ref{s:conclusions}.

\section{Data}
\label{s:data}

The data used here are taken from the \CNOCI\ cluster survey.
The observational procedures and data reduction are described in detail in
	\cite{CNOCtech}; only the  relevant features of the survey are discussed
	here.
The data were obtained using the multi-object spectrograph (MOS) at the
	Canada-France-Hawaii Telescope (CFHT).
A band-limiting filter was used with the spectrograph to reduce the length of
	each spectrum, further increasing the multiplexing rate.
Since the \CNOCI\ survey is a cluster redshift survey, each field was chosen
	to contain a cluster at its center -- most of the data is therefore
	unsuitable for investigating the clustering of field galaxies.
One field in the survey, MS1512+36, however, is well suited to a
	correlation analysis of field galaxies, since the cluster is quite poor.
Only data from this field are analyzed here; the complete data set is described
	and presented in \cite{CNOCclstr}.

Pencil-beam surveys present several problems related to the fairness of the
	sample (a sample is said to be fair if the structure contained within it is
	representative of the global average).
The first difficulty relates to the beam width; a typical beam diameter of
	$10\arcmin$ yields a width of $\order1.8\hmpc$ at $z=\slantfrac{1}{3}$
	($\Omega_0=1$), the same order of magnitude as the expected correlation
	length $r_0$ at that redshift, given any reasonable amount of evolution
	in the correlation function.

Thus, the distribution of objects in a beam may be dominated by a single large
	density inhomogeneity, biasing the estimation of $\xi$.
One possible solution to this problem is to calculate the correlation function
	from data taken from many pencil-beams scattered randomly throughout the
	sky.
This approach does not make optimal use of the data, however; it is preferable
	to place the beams so that the beam-beam separation is comparable to the
	beam width, thus increasing the number of pairs of objects with
	separations $\order r_0$.
The MS1512+36 data are from a mosaic of three fields from the survey, with a
	total angular size of $\dra\arcmin\times\ddec\arcmin$
	(222 square arcminutes). 
Despite the relatively large width of the sample ($4.9\hmpc$ at $z=0.36$
	if $\Omega_0=1$), there remains a large overdensity spanning the width of
	the field.
The effect of this inhomogeneity is discussed below.

A second, more subtle, problem with MOS data is that of selection effects.
Magnitude selection may bias the sample towards bright objects, which will
	lead to erroneous results if bright objects cluster differently from
	faint objects.
More importantly, MOS produces a lower limit on the separation of objects for
	which spectra may be observed.
Once one object is designated to be observed through a slit on a given mask,
	the placement of the spectra on the detector precludes designing another
	slit closely above or below the first.
This results in high density regions being sampled less completely than low
	density regions, thereby reducing the observed correlation.
Each of the three fields composing the MS1512+36 data was observed with 2
	different MOS masks, with a higher priority given in the second mask to
	completing observations of closely spaced pairs (see Yee \etal\ 
	1996 for details).
This somewhat reduces the amount of geometric selection.

To correct for the magnitude and geometric selection, the magnitude weight
	$w_m$ and local magnitude weight $w_{lm}$ are calculated from the data for
	each object in the sample.  
For a given object with apparent magnitude in the bin $(m,m+\Delta m)$ and
	observed redshift $z$, $w_m$ is proportional to the fraction of objects
	anywhere in the sample in the same magnitude bin which have observed
	redshifts.
The local magnitude weight $w_{lm}$ for an object is proportional to the
	fraction of objects in a circle with radius $120\arcsec$ about the first
	object in the same magnitude bin which have observed redshifts.
Also defined for each object is the geometric weight $w_{xy}=w_{lm}/w_m$,
	which is related to the number of nearby objects at any magnitude which
	have observed redshifts.
A detailed explanation of the weighting procedure is given in Yee \etal\
	(1996); a test of the extent to which these weights
	correct for the sampling nonuniformities is described in Section
	\ref{ss:corfunc}.

Three subsamples of the MS1512+36 data are created; the photometric, redshift
	and field samples.
The photometric sample, used for computing the angular correlation function,
	consists of the \nphot\ objects with $g-r$ colours, with Gunn $r$-band
	magnitude in the range $\mmin\le r\le\mmax$.
The upper limit is chosen so that the magnitude weight of every object in the
	redshift sample is less than $5$; the lower limit is employed since the
	masks were designed to exclude objects much brighter then the brightest
	cluster galaxy, which has $r=18.45$.

The redshift sample consists of all objects in the photometric sample which
	have identified redshifts in the range $\zmin\le z\le\zmax$.
These limits are chosen so that the spectral features used to identify
	emission line objects lie within the optimal response region of the filter.
This sample is shown in Figure \ref{f:slice}.
The use of band-limiting filters for the \CNOCI\ survey results in strong
	redshift selection effects in the sample, which are easily understood by
	noting the visibility of strong spectral features in our limited spectral
	window.
In this analysis we have extended the lower $z$ limit from 0.27 in
	\cite{CNOCtech}\ to 0.21.
The original higher limit was based on the detection of [\ion{O}{2}]
	$\lambda3727$ at the blue end of the spectrum.
However, the wavelength limits for the filter used are such that as
	[\ion{O}{2}] $\lambda3727$ disappears at the blue end, the [\ion{O}{3}]
	$\lambda\lambda4959,5007$ lines come in on the red end.
From our sample, we have found that whenever the [\ion{O}{3}]
	$\lambda\lambda4959,5007$ and [\ion{O}{2}] $\lambda3727$ lines are both
	within the spectral range, they are always detected simultaneously.
Hence we can safely use the [\ion{O}{3}] $\lambda\lambda4959,5007$ lines to
	extend our lower redshift limit; no significant selection bias as a
	function of spectral type is seen in the redshift sample (Yee \etal\
	1996).

\placefigure{f:slice}

The field sample is constructed from the redshift sample by removing the 
	\ndiff\ objects that have redshifts $0.3656\le z\le0.3796$.
This redshift range corresponds to a velocity range of $\pm2100\kms$, or six
	times the velocity dispersion of the cluster, at $z=0.3727$, the center of
	the cluster (Carlberg \etal\ 1996).
The field sample therefore almost certainly excludes all cluster members, thus
	removing the bias towards high density regions present in the redshift
	sample.
The field sample contains \nfield\ objects, with a median redshift of
	$\bar z=\zmed$.

As can be seen in Figure 1, the cluster appears to be embedded in an
	overdensity extending across the field.
In order to test the sensitivity of our results to the presence of this
	structure, we have computed the correlation functions for the field
	sample with all objects with $0.354\le z\le0.390$ removed.
The results are consistent with those computed from the entire field sample,
	although, as there are only 118 objects in the field sample outside the
	these extended redshift limits, the uncertainties are considerably larger.
We conclude that the presence of this structure does not unduly influence our
	results.

\section{Analysis}
\label{s:analysis}

\subsection{Estimating the Correlation Function}
\label{ss:corfunc}

The two-point correlation function $\xi$ is defined by (Peebles, 1980)
\begin{equation}
	\delta P(r) = \bar n\lp 1+\xi(r)\rp \delta V\;,
\label{eq:xir def}
\end{equation}
	where $\delta P(r)$ is the probability of finding a second object in a
	volume $\delta V$ with a physical separation $r$ from a randomly chosen
	object, and $\bar n$ is the mean density of objects.
For a finite sample of objects, and some (small) fixed separation difference
	$\Delta r$, $\xi(r)$ may be estimated from equation \eqref{xir def}\ as
\begin{equation}
	1+\xi(r)\approx{DD(r)\over N_D^2}{V\over\Delta V(r)}\;.
\label{eq:semi}
\end{equation}
	Here, $N_D$ is the number of objects in the sample, $DD(r)$ is the number
	of ordered pairs of objects in the sample with separation between $r$ and
	$r\!+\!\Delta r$, $V$ is the volume of the sample, and $\Delta V(r)$ is
	the average volume of the set surrounding a first object in which a second
	may be found with separation between $r$ and $r\!+\!\Delta r$ from the
	first.

The volumes $\Delta V$ in equation \eqref{semi}\ may be arbitrarily
	complicated; in practice, they are estimated using Monte-Carlo integration.
A random data set containing $N_R$ objects is generated within the volume of
	the original data set in a manner such that the random catalog is subject
	to the same selection criteria as the data.
If the number of pairs of objects with separation between $r$ and
	$r\!+\!\Delta r$, the first object belonging to the data set, the second
	to the random set, is $DR(r)$, then (Davis \& Peebles 1983)
\begin{equation}
	1+\xi(r)\approx{N_R\over N_D}{DD(r)\over DR(r)}\;.
\label{eq:xir est}
\end{equation}

The pair counts $DD$ and $DR$ in equation \eqref{xir est}\ may be computed
	with arbitrary weights, so that $DD(r) = \sum_{i,j}w_i^{(D)}w_j^{(D)}$
	and $DR(r) = \sum_{i,j}w_i^{(D)}w_j^{(R)}$, where the sums are taken over
	all data-data or data-random pairs of objects with separation between $r$
	and $r+\Delta r$, respectively, and $w_i^{(D)}$ and $w_i^{(R)}$ are the
	weights to be applied to $i$'th data and random object, respectively.
The object counts $N_D$ and $N_R$ in equation \eqref{xir est}\ are replaced
	with the weighted object counts $D = \sum_{i=1}^{N_D}w_i^{(D)}$ and
	$R = \sum_{i=1}^{N_R}w_i^{(R)}$, respectively.
The correlation function is then estimated as
\begin{equation}
	1+\xi(r) \approx {R\over D} {DD(r)\over DR(r)}\;.
\label{eq:xiestw}
\end{equation}
To correct for the selection effects present in this sample, and for the
	incompleteness present in any magnitude-limited sample, we take the total
	weight for an object at redshift $z$ with local magnitude weight $w_{lm}$
	to be $w_{tot} = w_{lm}/\phi(z)$, where $\phi(z)$ is the redshift
	selection function, defined as the fraction of objects at redshift $z$
	which lie within the apparent magnitude limits of the survey.
These total weights are then used for computing the pair and object counts
	defined above.  

The selected random catalog is generated by first creating a uniform random
	catalog, by randomly distributing objects throughout the sample volume in
	such a way that the comoving density of objects is constant.
Apparent magnitudes are then assigned to each object in the uniform random
	catalog using the process described below, and those objects lying outside
	the magnitude limits of the survey are discarded.
Finally, the local magnitude selection is estimated at the position of each
	object random object, and objects failing the selection criteria are
	discarded. 
For small data sets, this method is preferable to smoothing the observed
	redshift distribution, since the redshift distribution is dominated
	by density inhomogeneities comparable to the width of the sample, making
	the random distribution obtained sensitive to the smoothing window used.

Absolute magnitudes for the objects in the field sample are required to
	determine the luminosity function, which is used for generating
	the absolute magnitudes for the random objects.
The K-corrections are obtained by interpolating from model K-corrections in
	the $r$ and $g$ bands as a function of redshift for non-evolving galaxies
	of 4 spectral types (E+S0, Sbc, Scd, and Im).
The models are derived by convolving filter response functions with spectral
	energy distributions in Coleman, Wu, \& Weedman (1980).
These values are then corrected from the AB system to the standard Gunn system
	(Thuan \& Gunn 1976).
For each galaxy with redshift, a spectral classification is estimated by
	comparing the observed $g-r$ color with the model colors at the same
	redshift.
The spectral classification, obtained via interpolation, is treated as a
	continuous variable between the 4 spectral types.
From the spectral classification, the appropriate K-correction to the $r$
	magnitude is then derived using the models.

The $r$-band absolute magnitudes for the random catalog are generated
	according to the luminosity function for the data.
The luminosity function is modeled as a non-evolving Schechter function
	(Schechter, 1976)
\begin{equation}
	\Phi(M_r)=0.4\ln10\phi^\star10^{0.4(1+\alpha)(M_r^\star-M_r)}
		\exp\lp-10^{0.4(M_r^\star-M_r)}\rp\;,
\end{equation}
	where $\Phi(M_r)$ is the comoving number density of objects with absolute
	magnitude $M_r$ per unit magnitude.
The assumption of no evolution is reasonable, since the redshift range of the
	field sample is relatively small.
The parameters $M_r^\star$ and $\alpha$ are adjusted to fit the data using a
	least-squares fit.
The luminosity function (after an appropriate renormalization, which removes
	the dependence on $\phi^\star$) is then used as the probability
	distribution for the absolute magnitudes of the objects in the
	uniform random catalog.
An intrinsic colour $(g-r)_0$ is then chosen for each object, in a manner
	such that the distribution of intrinsic colours in the uniform
	random catalog is the same as that in the data catalog.
The $r$-band absolute magnitude $M_r$ and intrinsic colour $(g-r)_0$ for each
	object are then used to compute the K-corrections, using the method
	described in Section \ref{s:data}.

For each object in the uniform random catalog with apparent magnitude within
	the sample limits, magnitude and geometric weights $w_m$ and $w_{xy}$ are
	estimated by interpolating from the data weights.
The magnitude weight interpolation is a straightforward one-dimensional
	interpolation.
The two-dimensional geometric weight interpolation is performed by convolving
	the spatial map of geometric weights of each data object with a
	Gaussian with dispersion $\sigma=1\arcmin$.
The values of the convolved weights at the position of the random object are
	then summed, yielding a weight for that random object.
The local magnitude weight $w_{lm}$ for the random object is than calculated
	from $w_m$ and $w_{xy}$; the object is discarded unless $1/w_{lm}$ is less
	than a randomly chosen number between 0 and 1.
The redshift selection function $\phi(z)$ is then computed from the luminosity
	function parameters $M_r^\star$ and $\alpha$ and the sample magnitude
	limits; $w_{lm}$ and $\phi(z)$ are then combined to give a total weight
	$w_{tot} = w_{lm}/\phi(z)$ for each remaining object in the selected
	random catalog.

The redshift distribution for a random catalog generated in this manner is
	compared to the redshift distribution of the redshift sample in Figure
	\ref{f:dNdz}.
The random distribution has been rescaled to have the same number of objects
	within the redshift sample redshift range as the data distribution in this
	figure.\
Note that the random distribution is much more uniform than the distribution
	that would be obtained by smoothing the data distribution, due to the
	presence of the large structure discussed earlier.

\placefigure{f:dNdz}

This method of generating the random catalog and the weighting procedure
	are tested by comparing the angular correlation functions of the
	photometric and redshift samples.
The angular correlation function $w$ is estimated in a manner similar to that
	given in equation \eqref{xiestw};
\begin{equation}
	1+w(\theta)\approx{R\over D} {DD(\theta)\over DR(\theta)}\;,
\label{eq:west}
\end{equation}
	where the pair counts are now taken over pairs with angular separation
	$\theta$.
If the model \eqref{plawdef}\ for $\xi$ is correct, then (Peebles 1980)
\begin{equation}
	w(\theta) = A_w \lp\theta\over1''\rp^{-\delta}\;,
\label{eq:wplaw}
\end{equation}
	where $\delta=\gamma-1$.

The random catalog used for computing $w$ for the redshift sample is generated
	using the procedure described above, while the catalog for the photometric
	sample was generated uniformly; both catalogs contain $10\,000$ objects.
Figure \ref{f:angular}\ shows the results of a power-law fit to the angular
	correlation for the photometric sample, with $\delta$ fixed to \delfrc,
	and $w$ for the redshift sample. 
The amplitudes in equation \eqref{wplaw}\ are found to be  $A_w=\awpe$ for the
	photometric sample, and $A_w=\awce$ for the redshift sample.
The consistency of these amplitudes indicates that the total weights adequately
	describe the selection effects present in the data.
The photometric sample value of $A_w$ implies $w(1\arcmin)\approx0.07$,
	consistent with the value found by Infante \& Pritchet (1995) for a sample
	of objects with $F\leq22$.
The uncertainties in all correlations and fitted parameters given here are the
	$68.3\%$ confidence intervals computed using the bias-corrected bootstrap
	resampling method, described by \cite{EnF}.
Three hundred resamplings are used for each computation, the minimum
	recommended for this method.

\subsection{The Real Space Correlation Function}
\label{ss:realsp}

Two different correlation functions are derivable from redshift data; the
	redshift space correlation function $\xi(s)$ and the real space 
	function $\xi(r)$.
If the separation of a pair of objects is computed directly from redshifts, it
	includes the line-of-sight component of the object's peculiar
	velocities relative to the Hubble flow.
For the range of separations of interest here, the random internal motions of
	bound groups of objects dominate, elongating structures along the line of
	sight.
This elongation reduces the observed correlation at small separations, so the
	power-law model \eqref{plawdef}\ is expected to be valid only for the real
	space correlation function.

\subsubsection{Method}

Although it is impossible to measure real space separations directly using
	only redshift data, it is possible to estimate the parameters of a model
	for the real space correlation function.
This is accomplished by decomposing the redshift space separation of a pair of
	objects into components parallel and perpendicular to the line of sight to
	the pair.
Since the redshift space distortions act only along the line of sight,
	functions of the perpendicular component must be independent of these
	perturbations.

The decomposition is performed assuming the separations are small, so that
	the effects of curvature may be neglected.
Thus, given two objects with redshifts $z_1$ and $z_2$ and angular separation
	$\theta_{12}$, two vectors ${\bf x}_i\; (i=1,2)$ are formed, such that
\begin{equation}
	\lb {\bf x}_i\rb = {2c\over H_0}
		{\Omega_0z-(2-\Omega_0)(\sqrt{1+\Omega_0z}-1)\over\Omega_0^2(1+z)}
		\qquad(i=1,2)
\end{equation}
	($\Lambda=0$), and
\begin{equation}
	{\bf x}_1\cdot{\bf x}_2 = \lb {\bf x}_1\rb\lb{\bf x}_2\rb\cos\theta_{12}\;.
\end{equation}
The comoving redshift space separation of the pair is then
	${\bf s} \approx {\bf x}_2 - {\bf x}_1$, and the line of sight to the pair
	is $\bar{\bf x} = {1\over2} \lp{\bf x}_1 + {\bf x}_2\rp$.
The components of the physical separation parallel and perpendicular to the
	line of sight are then
\begin{equation}
	\pi = {{\bf s}\cdot\bar{\bf x}\over(1+\bar z)\lb{\bf\bar x}\rb}
\label{eq:pidef}
\end{equation}
	and
\begin{equation}
	r_p = {\sqrt{\lb\bf s\rb^2 - \pi^2}\over1+\bar z}\;,
\end{equation}
	where $\bar z = {1\over2} \lp z_1 + z_2\rp$.
The definition of the correlation function \eqref{xir def} is then
	generalized to describe the probability in excess of random of finding
	an object with redshift space separation $(r_p,\pi)$ from a randomly chosen
	object.
This is estimated as
\begin{equation}
	1+\xi(r_p,\pi) \approx {R\over D}{DD(r_p,\pi)\over DR(r_p,\pi)}\;,
\label{eq:xirppi est}
\end{equation}
	where $DD(r_p,\pi)$ and $DR(r_p,\pi)$ are the weighted number of data-data
	and data-random ordered pairs with separations $(r_p,\pi)$, respectively,
	and $D$ and $R$ are the weighted object counts defined in section
	\ref{ss:corfunc}.

Although $\xi(r_p,\pi)$ is affected by the redshift space distortions
	described earlier, the projected correlation function $w_p(r_p)$, defined
	by
\begin{equation}
	w_p(r_p) = \int_{-\infty}^{\infty}\xi(r_p,\pi)d\pi
\label{eq:wp def}
\end{equation}
	is not.  Thus,
\begin{equation}
	w_p(r_p) = \int_{-\infty}^{\infty}{\xi\lp\sqrt{r_p^2+x^2}\rp dx}\;,
\end{equation}
	where the integral is over the real space correlation function.  If the
	power law model \eqref{plawdef} is employed, then (Davis \& Peebles 1983)
\begin{equation}
	w_p(r_p) = \sqrt{\pi}
		{\Gamma\lp\delta\over2\rp\over\Gamma\lp\delta-1\over2\rp}
		r_0\lp r_p\over r_0\rp^{-\delta}\;.
\label{eq:wp plaw}
\end{equation}

The integral in equation \eqref{wp def} must be truncated at some $\pi_{max}$
	for any real data set.
Figure \ref{f:wp cut} shows the quantity $2\pi \bar nJ_p(1\hmpc,\pi)$ \vs\
	$\pi$ for the field sample, where
\begin{equation}
	J_p(r_p,\pi) = \int_{-\pi}^{\pi}
      	\int_0^{r_p} \xi(r_p',\pi') r_p' dr_p' d\pi'\;.
\label{eq:Jp}
\end{equation}
Given this definition, the quantity $2\pi \bar nJ_p(r_p,\pi)$ represents the
	mean number of objects in excess of random within the cylinder with radius
	$r_p$ and length $2\pi$ centered on an object in the sample.
This function is expected to increase with $\pi$ for small $\pi$,
	approaching some limiting value.
As can be seen, cutoffs less than $\order3\hmpc$ exclude real power in
	$\xi(r_p,\pi)$, while noise appears to be the primary contributor to the
	integral \eqref{Jp} for $\pi\gtrsim50\hmpc$.
We adopt $15\hmpc$ as the cutoff to be used in equation \eqref{wp def}; no
	significant change in the derived correlation length is observed when
	the cutoff is varied between $5\hmpc$ and $35\hmpc$.

\placefigure{f:wp cut}

\subsubsection{Results}

The projected correlation function $w_p(r_p)$ is calculated for the field
	sample using equation \eqref{xirppi est}, with the integral truncated at
	$15\hmpc$.
The random catalog is generated using the method described in Section
	\ref{ss:corfunc}; the catalog contains $200\,000$ objects with redshifts.
The pair counts $DD$ and $DR$ and object counts $D$ and $R$ are calculated
	using the total weights defined in section \ref{ss:corfunc}.
The results are shown in Figure \ref{f:realsp}; $r_0$ is determined for
	$\Omega_0=1$ and $0.2$ by fitting the model \eqref{wp plaw} to the data
	with $\delta$ fixed to \delfrc.
The correlation length is found to be $r_0=\rofe$ $(\Omega_0=1)$ or
	$r_0=\rooe$ $(\Omega_0=0.2)$, considerably smaller than the values found
	locally.

If it is assumed that the population in our sample at $z\order0.36$ will
	evolve to the population observed at $z=0$ in optical surveys, equation
	\eqref{ev}\ may be used, with $r_0(0)$ equal to the value observed in the
	local sample, to estimate $\epsilon$.
Applying equation \eqref{ev}, with $r_0=\rofe$ at $z=\zmed$ and $r_0=\roloc$
	at $z=0$ yields $\epsilon=\efe$ $(\Omega_0=1)$; using $r_0=\rooe$ gives
	$\epsilon=\eoe$ $(\Omega_0=0.2)$.
Thus, the correlation is found to be increasing with redshift in physical
	coordinates, although the $\Omega_0=0.2$ estimate of $\epsilon$ is
	consistent with no evolution.
Correlation fixed in comoving coordinates ($\epsilon=3-\gamma$) is effectively
	ruled out.

\placefigure{f:realsp}

An alternate explanation for the low correlation length is that some of the
	objects in our sample are weakly clustered and become intrinsically
	faint at the present epoch (Efstathiou \etal\ 1991).
That is, the evolution is in the luminosity function, not the correlation
	function. This is entirely feasible in our sample; if $r_0=\roloc$ locally,
	and $r_0=\roo$ at $z=\zmed$, and $\epsilon=0$, and the faint population is
	completely unclustered ($\xi=0$), then the fraction of objects in our
	sample which belong to this currently faint population is only
	$\order0.15$.
If the faint population is clustered realistically, say $r_0=3.8\hmpc$
	(as seen in the IRAS survey, Fisher \etal\ 1994a) for both the faint
	galaxy autocorrelation and faint-bright cross-correlation, then
	$\order40\%$ of the objects in our sample are required to
	be undetected in the present-epoch observations.
Thus, we cannot rule out evolution in the luminosity function as the source of
	the reduction in $r_0$ at intermediate redshifts.

\subsection{The Redshift Space Correlation Function}
\label{ss:redsp}

As noted in section \ref{ss:realsp}, the redshift space correlation function
	is not expected to be a power-law due to random peculiar velocities.
However, measurements of $\xi(s)$ can be used, in conjunction with a model for
	the real space correlation function, to provide information about the
	velocity distribution of objects in the sample.
This in turn yields information on the mean matter density.

\subsubsection{Method}

The real space correlation function $\xi(r)$ is related to $\xi(r_p,\pi)$ by
	(Peebles 1980)
\begin{equation}
	1+\xi(r_p,\pi) = \int g(r,{\bf v})\lp1+\xi(r)\rp d^3v\;,
\label{eq:3d}
\end{equation}
	where $g(r,{\bf v})$ is the distribution of relative peculiar pairwise
	velocities of pairs with separation $r$, and
	$r^2 = r_p^2 + \lp\pi-v_z/H(z)\rp^2$, where $v_z$ is the component of
	${\bf v}$ along the line of sight.
To obtain a relation between the redshift space correlation function $\xi(s)$
	and the real space correlation function $\xi(r)$, one integrates equation
	\eqref{3d} over a sphere of radius $s$;
\begin{equation}
	4\pi\int_{-s}^s\xi(s')s'^2ds'
		=2\pi\int\int_{-s}^s\int_0^{\sqrt{s^2-\pi^2}}\xi(r)
		g(r,{\bf v})r_rdr_pd\pi d^3v\;,
\label{eq:I1}
\end{equation}
	using the identity $2\pi\int_{-s}^s\int_0^{\sqrt{s^2-\pi^2}}\xi(r_p,\pi)
	r_pdr_pd\pi =4\pi\int_0^s\xi(s')s'^2ds'$.

We employ here a simplified model for the pairwise peculiar velocity
	distribution in which $g$ is independent of the separation $r$.
Two dimensions of the velocity integral in equation \eqref{I1} may therefore
	be performed immediately; we define the line-of-sight peculiar pairwise
	velocity distribution
	$f(v_z)=\int_{-\infty}^\infty\int_{-\infty}^\infty g(v_x,v_y,v_z)dv_xdv_y$.
Equations\eqref{I1} then reduces to
\begin{equation}
	4\pi\int_0^s\xi(s')s'^2ds'=
		2\pi\int_{-\infty}^\infty\int_{-s}^s\int_0^{\sqrt{s^2-\pi^2}}\xi(r)
		f(v_z)r_pdr_pd\pi dv_z\;,
\label{eq:I3}
\end{equation}
	where the integral on the left is over the redshift space correlation
	function, while the integral on the right is over the real space
	correlation function.
Differentiating equation \eqref{I3} with respect to $s$ gives the general
	relationship between the redshift space and real space correlation
	functions, under the assumption that $g$ is independent of $r$;
\begin{equation}
	\xi(s) = {1\over2}s^{-1}\int_{-\infty}^\infty\int_{-s}^s
		\xi \lp \sqrt{s^2-2 \lp v_z/H(z)\rp \pi +\lp v_z/H(z)\rp^2} \rp
	f(v_z)d\pi dv_z\;.
\label{eq:I4}
\end{equation}
The argument to the real space correlation function in equation \eqref{I4}
	is just the physical separation $r$, evaluated with $r_p=\sqrt{s^2-\pi^2}$.
If $\xi(r)$ is modeled as a power law \eqref{plawdef}, then the integral over
	$\pi$ in equation \eqref{I4} may be performed analytically, finally
	yielding
\begin{equation}
	\xi(s) = {1\over2(2-\gamma)}H(z)r_0^\gamma s^{-1}\int_{-\infty}^\infty
		\lp \lb s+v_z/H(z)\rb^{2-\gamma} - \lb s-v_z/H(z)\rb^{2-\gamma} \rp
		f(v_z) {dv_z\over v_z}\;.
\label{eq:xitheor}
\end{equation}

The simplified model for the velocity distribution used here takes $f$ to be an
	exponential with zero mean, and dispersion independent of separation;
\begin{equation}
	f(v_z) = \sqrt{1\over2\sigma^2}
		\exp\lp-\sqrt{2}\lb v_z\over\sigma\rb\rp\;.
\label{eq:vdist}
\end{equation}
Here, $\sigma^2$ is the projected pairwise peculiar velocity dispersion; the
	three-dimensional mean-square pairwise peculiar velocity
	$\avg{v^2}=3\sigma^2$, since the mean pairwise peculiar velocity is taken
	to be zero.

Given a value for the cosmological density parameter $\Omega_0$, $\sigma$
	may be estimated using the Cosmic Virial Theorem (Peebles 1980,
	Fisher \etal\ 1994b);
\begin{equation}
	\sigma^2(r,z) = {3 H(z)^2 \Omega(z) Q J r_0(z)^\gamma r^{2-\gamma} \over
		4b (\gamma-1)(2-\gamma)(4-\gamma)}\;,
\label{eq:CMV}
\end{equation}
	where $Q$ relates the two- and three-point correlation functions, $b$ is
	the linear bias factor and $J$ depends only on $\gamma$
	($J(\gamma=\gamfrc)=\Jfac$).
For $\gamma$ close to 2, $\sigma$ is almost independent of separation,
	consistent with equation \eqref{vdist}.
Equation \eqref{CMV} depends on the relation between the distributions of
	galaxies and matter through the (unknown) bias factor, and thus is of
	limited use as a probe of the true value of $\Omega_0$.
Note that according to this model, $\sigma$ evolves as
	$(1+z)^{-\epsilon/2}$ (holding $\gamma$ and $Q$ constant).

\subsubsection{Results}

\placefigure{f:redsp}

The redshift space correlation function is calculated for the field
	sample using the redshift space analogue of equation \eqref{xiestw}
\begin{equation}
	1+\xi(s)\approx{R\over D}{DD(s)\over DR(s)}\;,
\end{equation}
	where $DD(s)$ and $DR(s)$ are the number of data-data and data-random pairs
	with redshift space separations between $s$ and $s+\Delta s$, respectively.
The random catalog used here is the same as that used for computing the real
	space correlation function.
Figure \ref{f:redsp}\ shows $\xi(s)$ for the field sample, along with the
	predictions from equation \eqref{xitheor}, using $r_0=\rof$,
	$\gamma=\gamfrc$ and $\sigma=\sigf$ for $\Omega_0=1$, and $r_0=\roo$,
	$\gamma=\gamfrc$  and $\sigma=\sigo$ for $\Omega_0=0.2$
These values of $\sigma$ are computed using equation \eqref{CMV} with $Q=b=1$.
	Also shown are the curves given by equation \eqref{xitheor}\ using a
	Gaussian pairwise peculiar velocity distribution with $\sigma=140\kms$;
	this value corresponds to the mean velocity uncertainty in the sample of
	$100\kms$.

As can be seen, the $\sigma=\sigf$, $\Omega_0=1$ model overestimates the
	redshift space perturbations.
A least-squares fit of equation \eqref{xitheor} to the data, using equations
	\eqref{vdist} and \eqref{CMV}, with $\Omega_0=1$, $b=1$ and
	$\gamma=\gamfrc$ yields $r_0=\xirofe$, inconsistent with
	the value derived from the projected correlation function data,
	$r_0=\rofe$, with 90\% confidence.
The $\sigma=\sigo$, $\Omega_0=0.2$ model matches the data more closely,
	consistent with the low $\Omega_0$ favored by \cite{CfA}\ and \cite{IRAS2}.
However, the observed $\xi(s)$ is consistent with a model with zero
	pairwise peculiar velocity dispersion, and the redshift space distortions
	due solely to velocity measurement errors; more data are therefore needed 
	for a precise determination of $\sigma$.
We conclude that the data are best modeled by a low density parameter;
	taking $\Omega_0=0.2$, $r_0=\rooe$ and $\sigma=\sigo$ yields models which
	are consistent with both the observed $w_p(r_p)$ and $\xi(s)$.

\placefigure{f:redsp}

\section{Conclusions}
\label{s:conclusions}

We have found that the physical correlation length for $\zmin\le z\le\zmax$ is
	$r_0=\rofe$ if $\Omega_0=1$, implying $\epsilon=\efe$.
If $\Omega_0=0.2$, $r_0=\rooe$ and $\epsilon=\eoe$; the uncertainties are
	estimated using the bias-corrected bootstrap resampling method, with 300
	resamplings.
These results are consistent with earlier results obtained from angular
	surveys, which indicate rapid evolution (Efstathiou \etal\ 1991, Infante
	\& Pritchet 1995).
It is also consistent with the results from the CFRS (Le F\`evre \etal\, 1996).
This decrease in $r_0$ from its present value may be interpreted either
	as a real change in the clustering of the observed galaxies, or as due
	to a weakly clustered population at $z\order\zmed$ which is intrinsically
	faint at the present epoch.

The projected pairwise peculiar velocity dispersion at $z=\zmed$,
	$\sigma=\sigf$, predicted by the Cosmic Virial Theorem using
	$\Omega_0=1$ is inconsistent at the 90\% confidence level
	with the observed redshift space correlation function.
The $\Omega_0=1$ is therefore weakly rejected.
The $\Omega_0=0.2$ prediction, $\sigma=\sigo$, however, matches the
	data more closely.
Thus, the relatively small redshift space distortions present favour low
	$\Omega_0$ as determined from the Cosmic Virial Theorem, consistent with
	the results of \cite{CfA}\ and \cite{IRAS2}.

More data is required to obtain a precise value for $\sigma$, and larger
	scales need to be sampled in order to obtain a smooth redshift
	distribution, thereby removing uncertainties in $\xi$ due to density
	inhomogeneities on the scale of the sample diameter.
A larger sample would also enable computations of the correlation function for
	subsamples based on galaxy colour or intrinsic brightness, which would
	help distinguish between the two possible sources of observed evolution
	described in section \ref{ss:realsp}.
The \CNOCII\ redshift survey, presently in progress, will yield $\order5000$
	high-accuracy redshifts in the range $0.15\leq z\leq0.7$.
This survey will contain enough objects, and sample sufficiently large scales,
	to permit accurate computations of the redshift space and real space
	correlation functions and their evolution at intermediate redshifts.

We thank all participants of the \CNOCI\ cluster survey for assistance in
	obtaining and reducing these data.
The Canadian Time Assignment Committee for the CFHT generously allocated
	substantial grants of observing time, and the CFHT organization provided
	the technical support which made these observations feasible.
We gratefully acknowledge financial support from NSERC and NRC of Canada.

\newpage

\figcaption[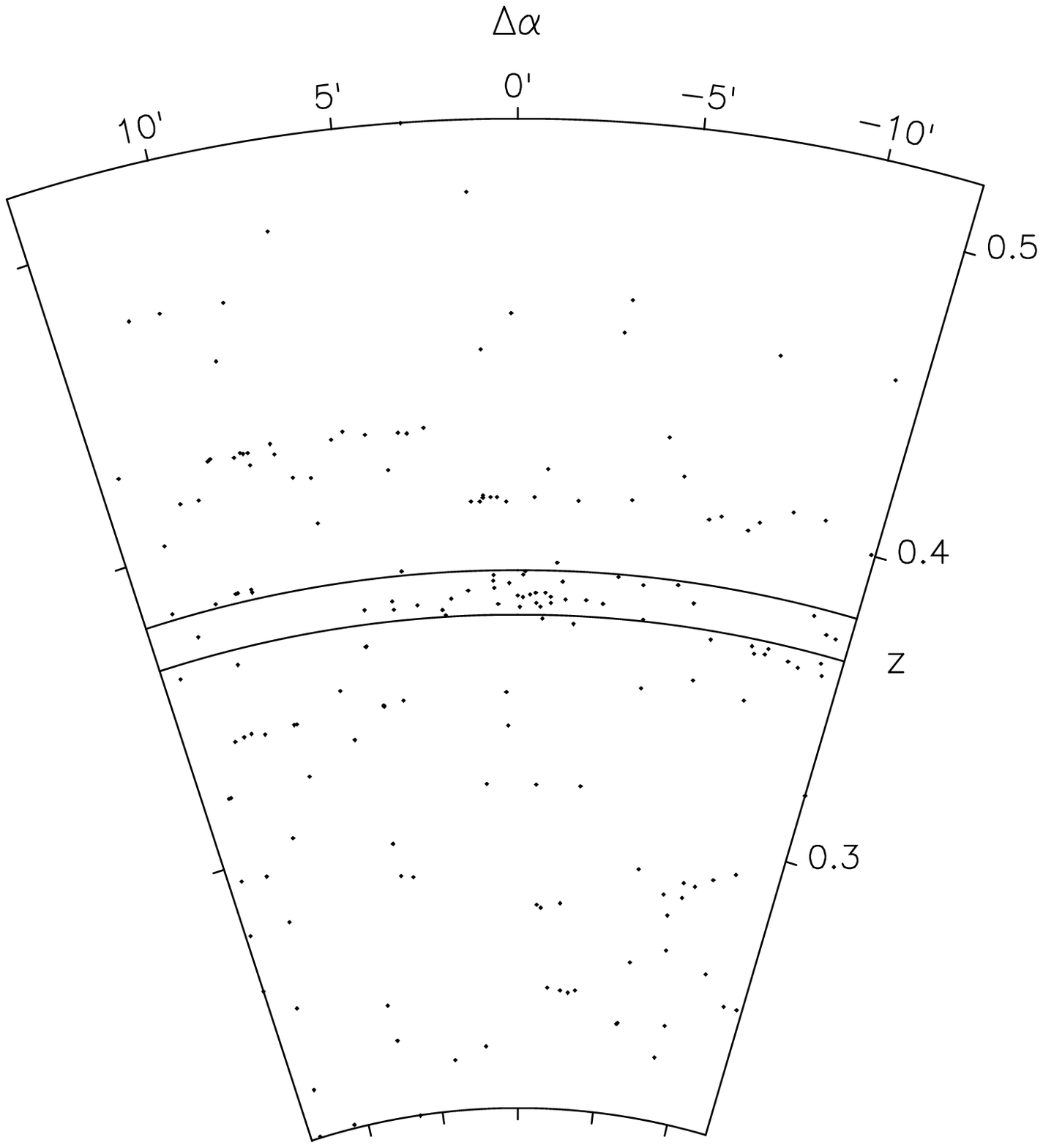]{
The MS1512+36 redshift sample.  The angular scale
has been expanded by a factor of $\order80$.  The objects within the central
box are not included in the field sample.
\label{f:slice}
}

\figcaption[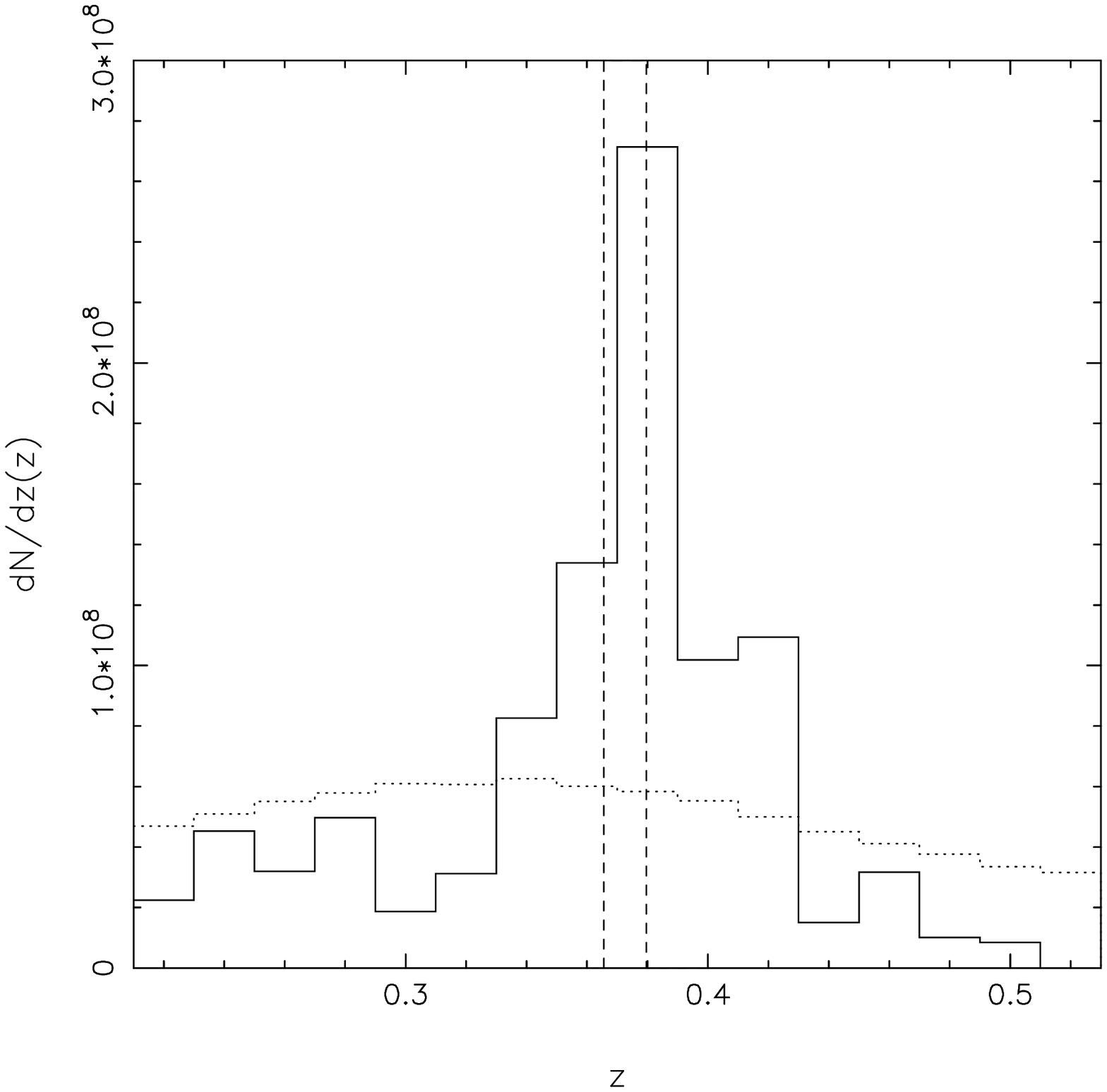]{
The redshift distribution ${dN\over dz}(z)$ for the redshift sample and
corresponding random catalog.  The solid line is ${dN\over dz}$ for the
redshift sample; the distribution of the random catalog is indicated by the
dotted line.  The two vertical lines indicate the redshift range
containing the \ndiff\ objects present in the redshift sample but not in the
field sample.  The random distribution has been normalized so as to have the
same integral as the data distribution over the redshift range shown.
\label{f:dNdz}
}

\figcaption[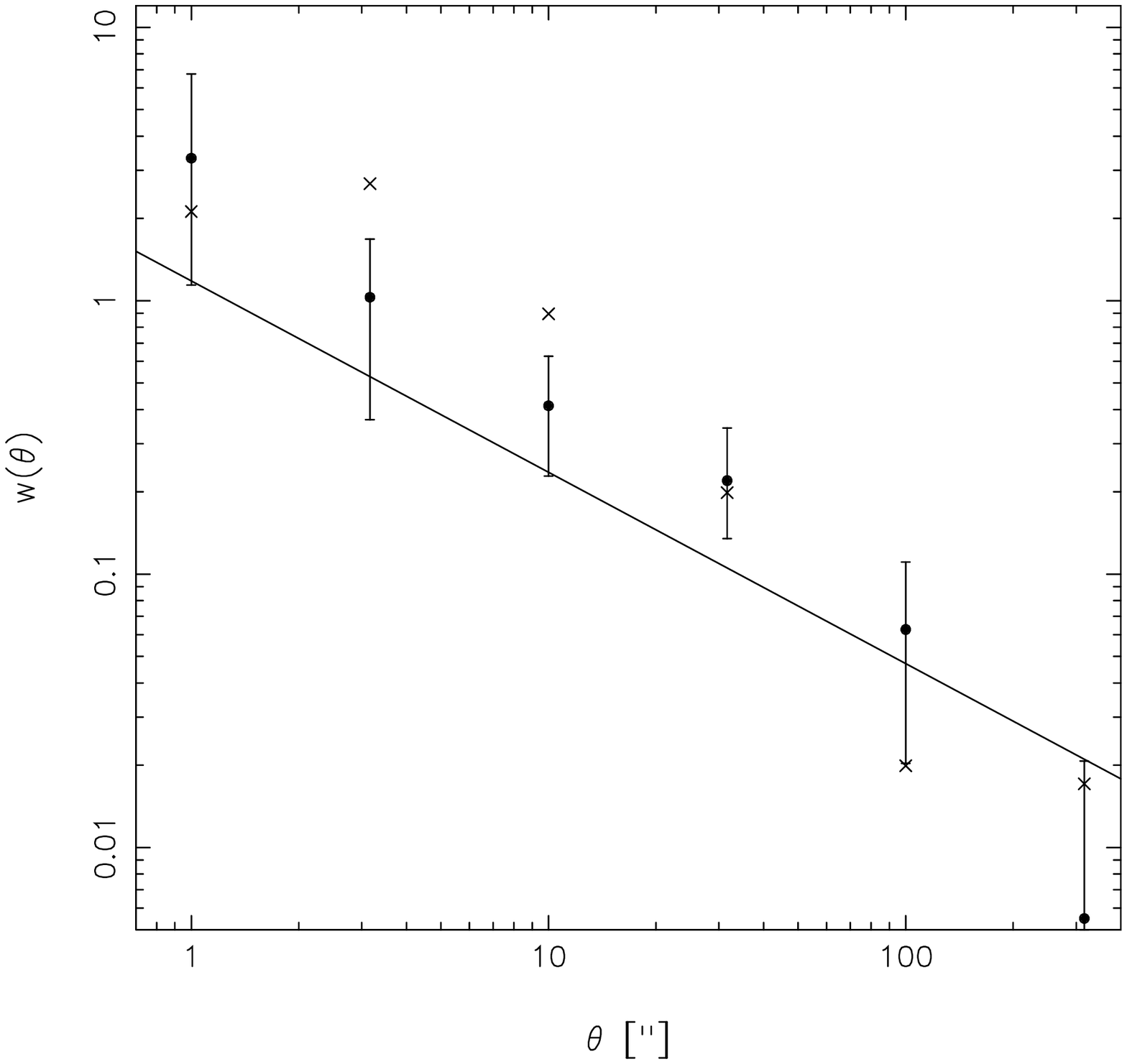]{
The angular correlation function $w(\theta)$ for the photometric and redshift
samples.  The values of $w$ for the photometric sample are indicated by dots;
the data for the redshift sample are indicated by crosses. The
error bars are the 68.3\% uncertainties estimated from 300 bootstrap
resamplings of the data.  The solid line is given by the least-squares fit to
the photometric data, with $\delta$ in equation (9) fixed
to \delfrc.
\label{f:angular}
}

\figcaption[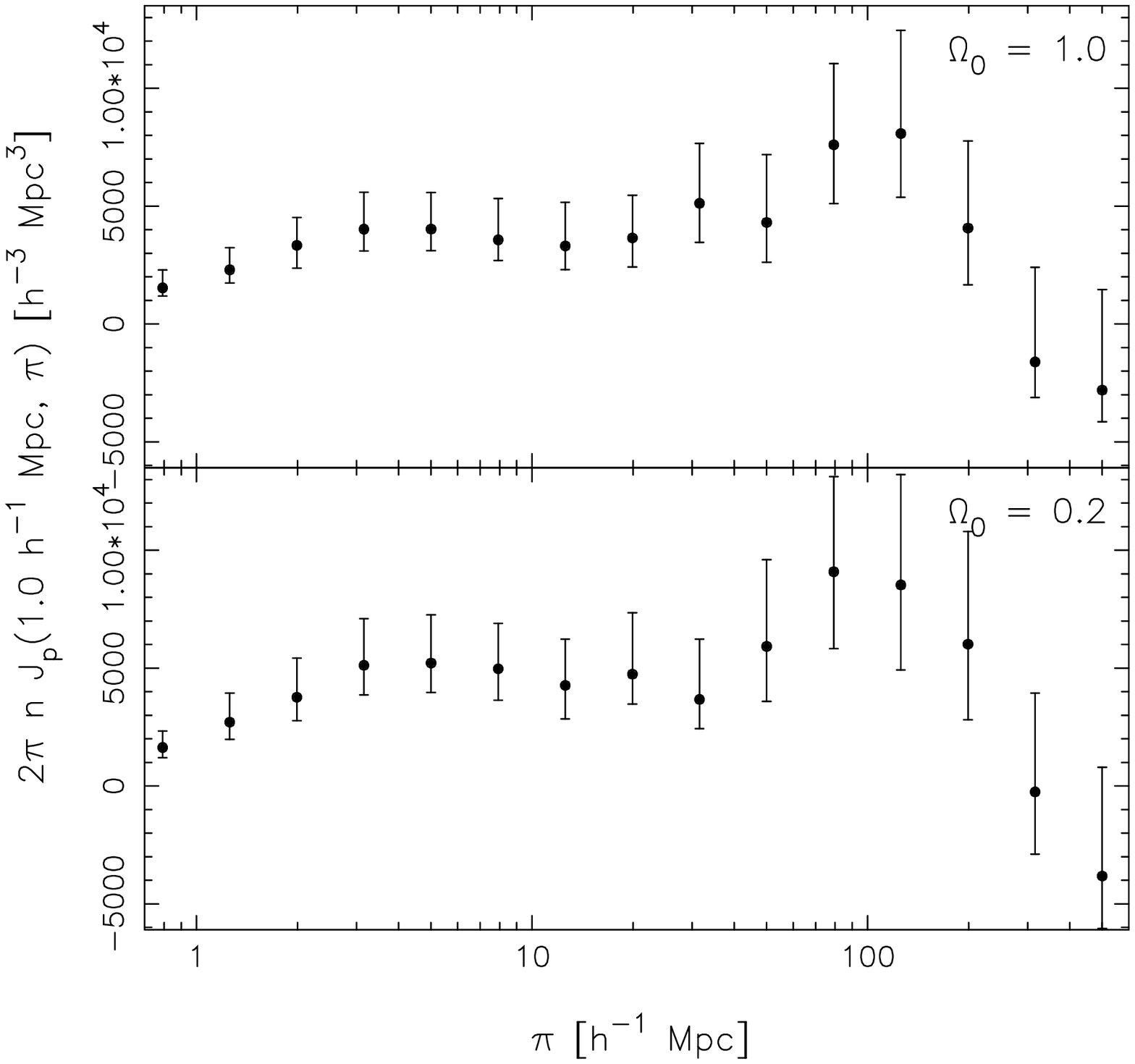]{
The counts in excess of random within a cylinder of radius $1\hmpc$ and
length $2\pi$, $2\pi\bar n$ $J_p(1\hmpc,\pi)$, for the field sample,
for $\Omega_0=1$ and
$0.2$.  The error bars are the 68.3\% bootstrap confidence intervals.
\label{f:wp cut}
}

\figcaption[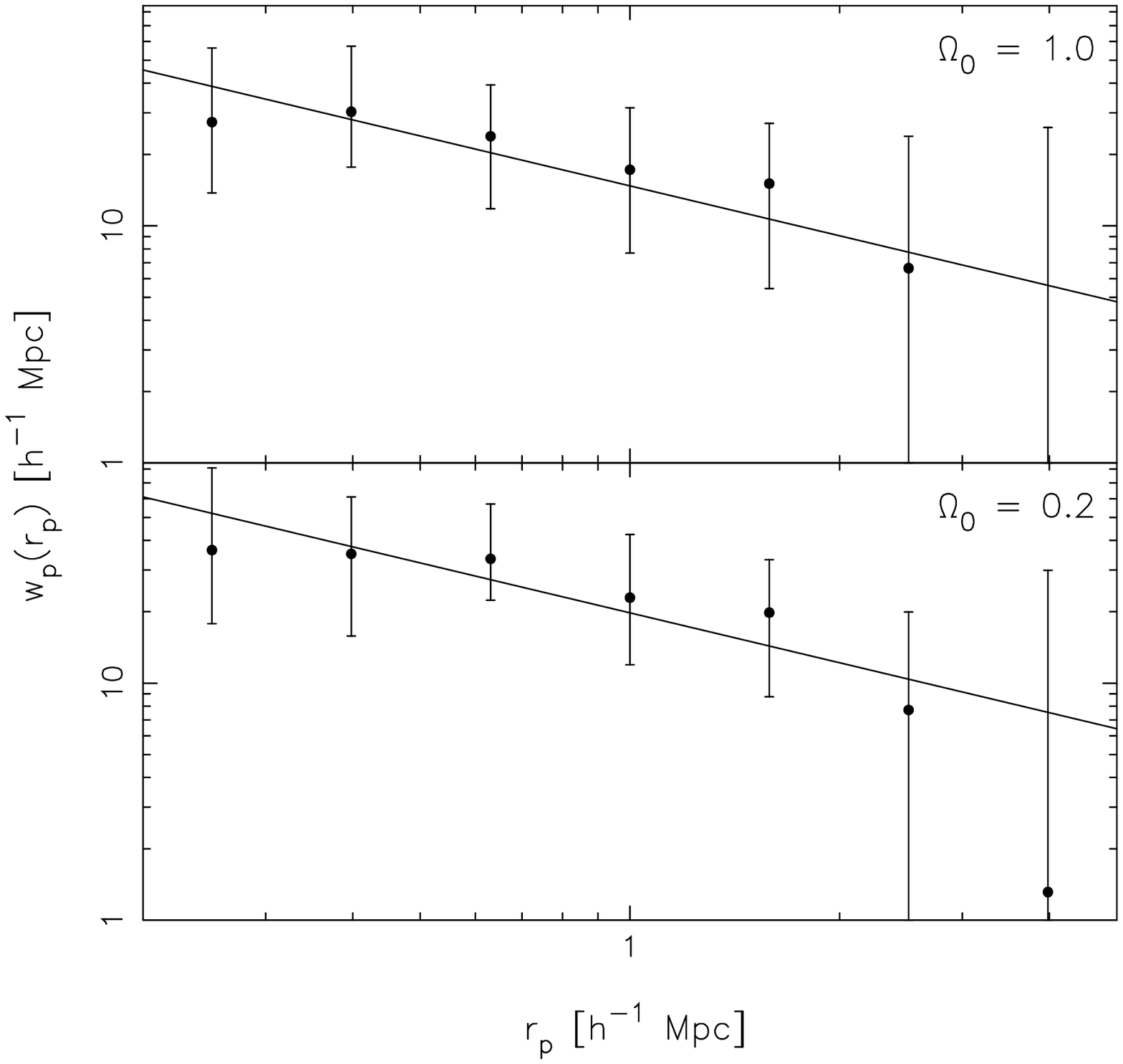]{
The projected correlation function $w_p(r_p)$ for the field sample, for
$\Omega_0=1$ and $0.2$.  The error bars are the 68.3\% bootstrap
confidence intervals.  The solid line in each panel is from the least-squares
fit of equation (17) to the data with $\delta$ fixed to \delfrc.
\label{f:realsp}
}

\figcaption[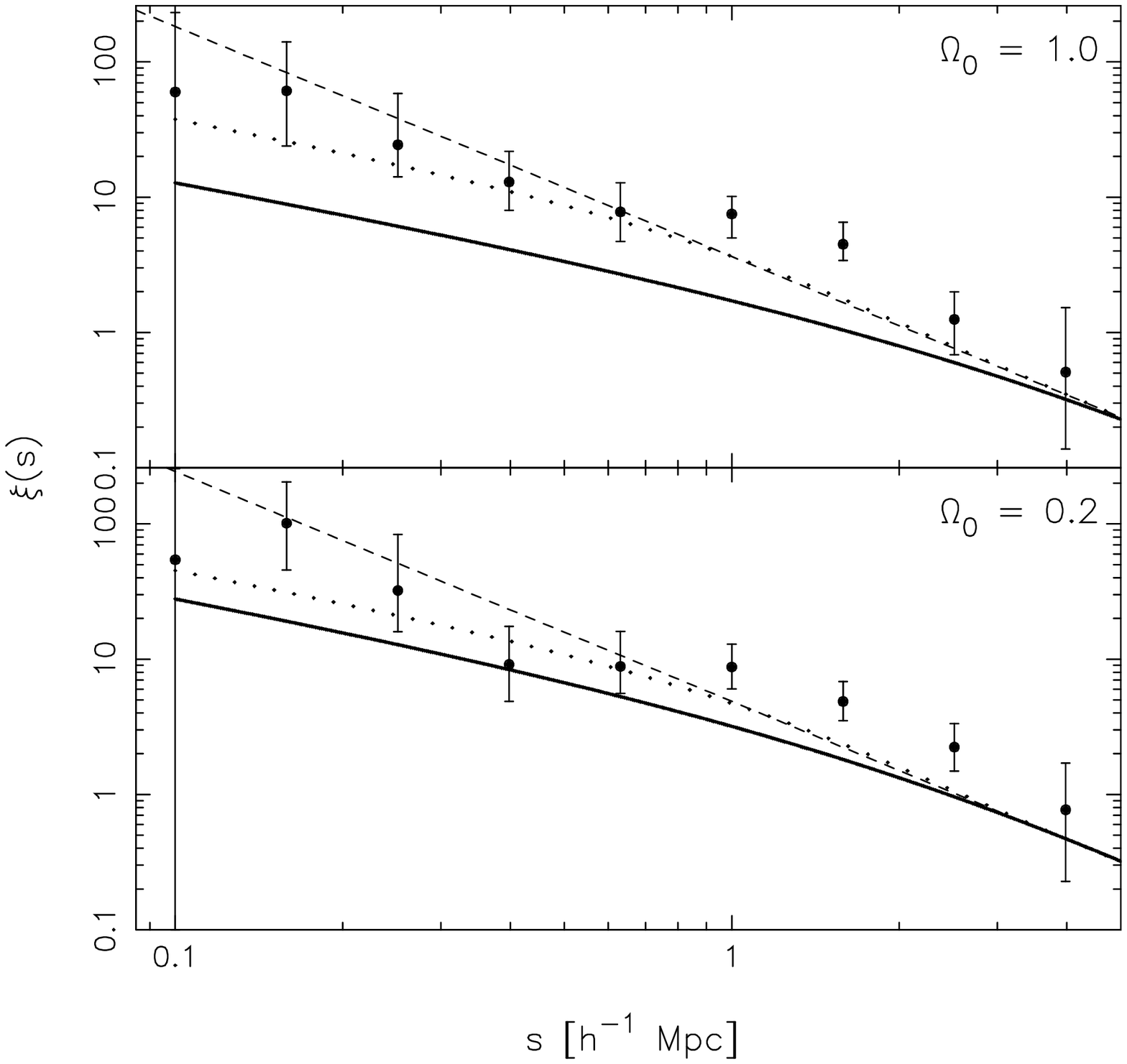]{
The redshift space correlation function $\xi(s)$ for the field sample, for
$\Omega_0=1$ and $0.2$.  The error bars are the 68.3\% bootstrap
confidence intervals.  The dashed line in each panel is the a power-law
with $\gamma=\gamfrc$ and $r_0$ taken from the
fit to the corresponding projected correlation function data.  The solid
curve in each panel is given by equation (23), with
$\sigma=\sigf$ ($\Omega_0=1$) and $\sigo$ ($\Omega_0=0.2$).  The
dotted curve is obtained using a Gaussian with $\sigma=140\kms$ in
place of the exponential in the velocity distribution model (24).
\label{f:redsp}
}

\begin{figure}
\figurenum{1}
\plotone{figure1.ps}
\caption{}
\end{figure}
\begin{figure}
\figurenum{2}
\plotone{figure2.ps}
\caption{}
\end{figure}
\begin{figure}
\figurenum{3}
\plotone{figure3.ps}
\caption{}
\end{figure}
\begin{figure}
\figurenum{4}
\plotone{figure4.ps}
\caption{}
\end{figure}
\begin{figure}
\figurenum{5}
\plotone{figure5.ps}
\caption{}
\end{figure}
\begin{figure}
\figurenum{6}
\plotone{figure6.ps}
\caption{}
\end{figure}

\end{document}